\documentclass[aps,pra,showpacs,superscriptaddress,preprint]{revtex4}
\usepackage{amssymb}
\usepackage{amsmath}
\usepackage{graphicx}

\catcode`ð=\active
 \defð{\u{g}}
 \catcode`Ð=\active
 \defÐ{\u{G}}
 \catcode`Ý=\active
\defÝ{\. I}
 \catcode`ö=\active
\defö{\"{o}}
 \catcode`Ö=\active
 \defÖ{\"O}
 \catcode`ü=\active
 \defü{\"{u}}
 \catcode`Ü=\active
 \defÜ{\"{U}}
 \catcode`Þ=\active
\defÞ{\c{S}}
 \catcode`þ=\active
 \defþ{\c{s}}
 \catcode`ý=\active
 \defý{{\i}}
 \catcode`ç=\active
\defç{\d{c}}
 \catcode`Ç=\active
\defÇ{\d{C}}

\input{tcilatex}

\begin{document}

\title{Spectra of cylindrical quantum dots: the effect of electrical and
magnetic fields together with AB flux field }
\author{Sameer M. Ikhdair}
\email[E-mail: ]{sikhdair@neu.edu.tr}
\affiliation{Physics Department, Near East University, 922022, Nicosia, North Cyprus,
Mersin 10, Turkey}
\author{Majid Hamzavi}
\email[E-mail: ]{majid.hamzavi@gmail.com (Corresponding author)\\
Tel.:+98 273 3395270, fax: +98 273 3395270}
\affiliation{Department of Basic Sciences, Shahrood Branch, Islamic Azad University,
Shahrood, Iran}
\author{Ramazan Sever}
\email[E-mail: ]{sever@metu.edu.tr}
\affiliation{Physics Department, Middle East Technical University, 06531, Ankara, Turkey}
\date{%
\today%
}

\begin{abstract}
We study the spectral properties of electron quantum dots (QDs) confined in
2D parabolic harmonic oscillator influenced by external uniform electrical
and magnetic fields together with an Aharonov-Bohm (AB) flux field. We use
the Nikiforov-Uvarov method in our calculations. Exact solutions for the
energy levels and normalized wave functions are obtained for this exactly
soluble quantum system. Based on the computed one-particle energetic
spectrum and wave functions, the interband optical absorption GaAs spherical
shape parabolic QDs is studied theoretically and the total optical
absorption coefficient is calculated.

Keywords: Parabolic harmonic oscillator, Quantum dots, Electrical and
magnetic fields, AB flux field, Light interband transition, Threshold
frequency of absorption
\end{abstract}

\pacs{03.65.-w; 03.65.Fd; 03.65.Ge; 71.20.Nr; 73.61.Ey; 73.63.Kv; 85.35.Be}
\maketitle

\newpage

\section{Introduction}

In the recent years, the subject of quantum dots (QDs) as low-dimensional
quantum systems have been the focus of extensive theoretical investigations.
Much efforts have recently been done into understanding their electronic,
optical and magnetic properties. The application of magnetic field is
equivalent to introducing an additional confining potential which modifies
the transport and optical properties of conduction-band electrons in QDs. In
addition, introducing electric field gives rise to electron redistribution
that makes change to the energy of quantum states which experimentally
control and modulate the intensity of optoelectronic devices [1,2]. Indeed,
it is worthwhile to investigate the influence of electric and magnetic
fields on the electrons in QDs. Experimental research is currently made to
investigate the nonlinear optical and quantum properties of low-dimensional
semiconducting structures for the fabrication purposes and subsequent
working of electronic and optical devices [3-14]. A number of works takes
the effects of an electric or a magnetic field into account in studying
quantum wells, quantum wires and QDs [9-14]. For practical and theoretical
reasons, more works analyzing these structures have been focused on the
interband light absorption coefficient and magnetic properties with
restricted geometries [15] of spherical [16,17,18], parabolic, cylindrical
and rectangular [19]\ QDs and other nanostructures such as superlatices,
quantum wires, wells, antidots, well wires and antiwells [20,21,22] in the
presence and absence of magnetic field [1,2]. In recent years, the rapid
development in semiconductor physics and in nanostructures technology
provides wide techniques for the possibility of fabrication of
low-dimensional quantum structures like quantum wells, quantum wires and
quantum dots which can be treated with high accuracy as two-,one- and
zero-dimensional nanostructures, respectively [23].

Harmonic oscillator belongs to the most important and most commonly used
physical models. Due to the formal simplicity, it is considered as one of
the exactly solvable quantum mechanical problems. It is used to model a wide
variety of phenomena ranging from molecular vibrations to the behaviour of
quantized fields. The Schr\"{o}dinger equation for an electron in a uniform
magnetic field confined by a harmonic oscillator type potential was solved
in 1928 by Fock [24] and Darwin [25]. There are many recent studies on $n$%
-particle systems confined in a nonrelativistic harmonic oscillator
potential [26]\ and rotation-vibration spectra of diatomic molecules [27].
Harmonic oscillator potential may be used to describe spatial confinement of
quantum objects, the effects of embedding particles in nano-cavities, in
fullerenes, in liquid helium [28,29].

A relativistic harmonic oscillator is far from being trivial and is not
unique. Nikolsky [30] and Postepska [31] studied Dirac equation for an
electron in the field of a quadratic potential. The eigenvalue problem
reduces to a quartic equation with no bound solutions. Toyama and Nogami
[32] discussed the relativistic systems which have infinite number of bound
states whose energy states are all equally spaced using the inverse
scattering method [33]. An approach leading to the Dirac oscillator based on
construction of exactly solvable Dirac equation which in the
non-relativistic limit reduces to Schr\"{o}dinger harmonic oscillator
equation [34,35].

Recently, interband optical absorption in GaAs spherical shape parabolic QDs
in the presence of electrical and magnetic fields was investigated by Atoyan
et al [1,2]. They solved the Schr\"{o}dinger equation for a spinless
particle confined by a two-dimensional (2D) cylindrical harmonic oscillator
potential. Su and Ma [36] solved the 3D and 1D Dirac equations with both
scalar and vector harmonic oscillator potentials. Qiang [37] obtained the
bound state energies and normalized wave functions for the Klein-Gordon (KG)
and Dirac equations with equal scalar and vector harmonic oscillator
potentials. The 2D quantum systems can probe the connection between
classical and quantum chaos and has application in a number of surface
systems such as atomic corrals [38]. For this reason, Qiang [39] studied the
energy formulas and their corresponding normalized wave functions of a 2D
relativistic quantum harmonic oscillator system for the first time. In
particular, Qiang solved the spinless (spin-$0)$ KG and spin-$1/2$ Dirac
equations with equal scalar and vector harmonic oscillator potentials in 2D
space and obtained the normalized wave functions and formulas for energy.

Very recently, we have studied the exact analytical bound state energy
eigenvalues and normalized wave functions of the spinless relativistic
equation with equal scalar and vector pseudoharmonic interaction under the
effect of external uniform magnetic field and AB flux field [40] in the
framework of the Nikiforov-Uvarov (NU) method [41,42,43,44]. The
non-relativistic limit of our solution is obtained by making an appropriate
mapping of parameters. Further, the KG-pseudoharmonic and KG-harmonic
oscillator special cases are also treated. Furthermore, we carried out
detailed exact energetic spectrum and wave functions of the Schr\"{o}dinger
equation with a pseudoharmonic potential in the presence of external
magnetic and AB flux fields [45]. The low-lying energy levels serve as a
base for calculating the corresponding interband light (optical) absorption
coefficient and the threshold frequency value of absorption for the given
model. In addition, the effect of the temperature on the effective mass is
also calculated for GaAs semiconductor.

An attempt is made in this paper to investigate the Schr\"{o}dinger equation
describing a spinless particle confined by a 2D parabolic harmonic
oscillator potential when external uniform electrical and magnetic fields
are applied together with Aharonov-Bohm (AB) flux field. We obtain the
energy spectra and wave functions in the non-relativistic harmonic
oscillators. So, we solve the Schr\"{o}dinger equation in Refs. [1,2] in the
presence of AB flux field too. The NU method [41,42,43,44] is used in the
present solution. In addition, the interband light absorption coefficient
and the threshold frequency value of absorption are calculated.

The structure of the paper is organized as follows. In Sec. 2, we
investigate the Schr\"{o}dinger particle in QDs confinement 2D parabolic
harmonic oscillator potential when electrical and magnetic fields together
with Aharonov-Bohm (AB) flux fields are applied in the framework of the NU
method. The exact analytical expressions for the energy formulas and
normalized wave functions are calculated. We also calculate the direct
interband light absorption coefficient and the threshold frequency of
absorption. The paper ends with a brief concluding remarks in Sec. 3.

\section{Theory and Calculations}

In this section, we shall consider the solution of spinless Schr\"{o}dinger
equation for the harmonic oscillator potential influenced by electrical,
magnetic and AB flux fields. The NU method [41,42,43,44] which has been
proved its success is used in our treatment.

\subsection{Bound-state solutions of the 2D harmonic oscillators}

Consider a two-dimensional ($2D$) single charged electron, $e,$ with an
electronic effective mass $\mu $ (for GaAs, $\mu =0.067m_{0}$) in the
conduction band, confined to a parabolic potential like quantum dots (QDs).
We will study the spectral properties with QDs confinement parabolic
harmonic oscillator potential influenced by uniform electrical and magnetic
fields together with an Aharonov-Bohm (AB) flux field, applied
simultanously. In cylindrical coordinates, the Schr\"{o}dinger equation
describing a spinless (spin-$0$) electron in such a quantum system is
usually written in the form [46]%
\begin{equation}
\left[ \frac{1}{2\mu }\left( \overrightarrow{p}+\frac{e}{c}\overrightarrow{A}%
\right) ^{2}-e\overrightarrow{\mathcal{E}}\cdot \overrightarrow{z}+V_{\text{%
conf}}(\vec{r})\right] \psi (\rho ,\varphi ,z)=E\psi (\rho ,\varphi ,z),
\end{equation}%
where $\overrightarrow{p}$ is the vector momentum with the magnetic field $%
\overrightarrow{H}=\overrightarrow{\nabla }\times \overrightarrow{A}$ (in
the symmetric gauge vector potential $\overrightarrow{A}=(A_{\rho
}=A_{z}=0,A_{\varphi }=H\rho /2)$), $\overrightarrow{\mathcal{E}}$ =$%
\mathcal{E}\widehat{z}$ is the applied electrostatic field along the $z$
axis and the QDs confinement parabolic potential is taken as [1,2] 
\begin{equation}
V_{\text{conf}}(\rho ,z)=\frac{1}{2}\mu \omega ^{2}r^{2}=\frac{1}{2}\mu
\omega ^{2}\left( \rho ^{2}+z^{2}\right) ,
\end{equation}%
where $\omega $ is the frequency of QDs measuring the strength of
confinement potential given by%
\begin{equation}
\omega \sim \frac{\hbar }{\mu r_{0}^{2}},
\end{equation}%
and $r_{0}$ is the oscillator length. The vector potential $\overrightarrow{A%
}$ of the magnetic field may be represented as a sum of two terms, $%
\overrightarrow{A}=\overrightarrow{A}_{1}+\overrightarrow{A}_{2}$ such that $%
\overrightarrow{\nabla }\times \overrightarrow{A}_{1}=\overrightarrow{H}$
and $\overrightarrow{\nabla }\times \overrightarrow{A}_{2}=0,$ where $%
\overrightarrow{H}$ $=H\widehat{z}$ is the applied magnetic field pointing
in the positive $z$ direction and is thus parallel to the two plane-parallel
electrodes of infinite extent, and $\overrightarrow{A}_{2}$ describes the
additional magnetic flux $\Phi _{AB}$ created by a solenoid inserted inside
the QDs. Hence, the vector potentials have azimuthal components, in the
cylindrical coordinate system, given by [40,45,47,48] 
\begin{equation*}
\overrightarrow{A}_{1}=\left( A_{1\rho }=0,\text{ }A_{1\varphi }=\frac{H\rho 
}{2},\text{ }A_{1z}=0\right) ,\text{ }\overrightarrow{A}_{2}=\left( A_{2\rho
}=0,\text{ }A_{2\varphi }=\frac{\Phi _{AB}}{2\pi \rho },\text{ }%
A_{2z}=0\right) ,
\end{equation*}%
\begin{equation}
\overrightarrow{A}=\overrightarrow{A}_{1}+\overrightarrow{A}_{2}=\left(
A_{\rho }=0,\text{ }A_{\varphi }=\frac{H\rho }{2}+\frac{\Phi _{AB}}{2\pi
\rho },\text{ }A_{z}=0\right) .
\end{equation}
The Schr\"{o}dinger equation (1) with potential (2) in cylindrical
coordinates has a form%
\begin{equation*}
-\frac{\hbar ^{2}}{2\mu }\left( \frac{1}{\rho }\frac{\partial }{\partial
\rho }\left( \rho \frac{\partial }{\partial \rho }\right) +\frac{1}{\rho ^{2}%
}\frac{\partial ^{2}}{\partial \varphi ^{2}}+\frac{\partial ^{2}}{\partial
z^{2}}\right) \psi (\rho ,\varphi ,z)-\left( \frac{i\hbar \omega _{c}}{2}+%
\frac{i\hbar e\Phi _{AB}}{2\pi \mu c}\frac{1}{\rho ^{2}}\right) \frac{%
\partial \psi (\rho ,\varphi ,z)}{\partial \varphi }
\end{equation*}%
\begin{equation*}
+\left( \frac{\mu (\omega _{c}^{2}+4\omega ^{2})}{8}\rho ^{2}+\frac{\mu
\omega ^{2}}{2}z^{2}+\frac{e^{2}\Phi _{AB}^{2}}{8\pi ^{2}\mu c^{2}}\frac{1}{%
\rho ^{2}}+\frac{e^{2}H\Phi _{AB}}{4\pi \mu c^{2}}-e\mathcal{E}z\right) \psi
(\rho ,\varphi ,z)=E\psi (\rho ,\varphi ,z),
\end{equation*}%
\begin{equation}
E=E_{\rho }+E_{z},
\end{equation}%
where $\psi =\psi (\rho ,\varphi ,z)$ is a wave function and $\omega
_{c}=eH/\mu c$ is the cyclotron frequency. $.$

Let us take the wave function ansatz for an electron as 
\begin{equation}
\psi (\rho ,\varphi ,z)=R(\rho ,\varphi )\chi (z),\text{ }R(\rho ,\varphi
)=g(\rho )e^{im\varphi },\text{ }m=0,\pm 1,\pm 2,\ldots ,
\end{equation}%
where $m$ is the magnetic quantum number. Upon inserting the above wave
function into Eq. (5), we shall obtain equations whose solutions are $g(\rho
)$ and $\chi (z)$ [1,2,49]:%
\begin{equation}
g^{\prime \prime }(\rho )+\frac{1}{\rho }g^{\prime }(\rho )+\left( \frac{%
2\mu E_{\rho }}{\hbar ^{2}}-\frac{\mu \omega _{c}}{\hbar }(m+\alpha )-\frac{%
\mu ^{2}\Omega ^{2}}{4\hbar ^{2}}\rho ^{2}-\frac{(m+\alpha )^{2}}{\rho ^{2}}%
\right) g(\rho )=0,
\end{equation}%
with%
\begin{equation}
\Omega =\sqrt{\omega _{c}^{2}+4\omega ^{2}},\text{ }\alpha =\frac{\Phi _{AB}%
}{\Phi _{0}},\text{ }\Phi _{0}=\frac{hc}{e},
\end{equation}%
where $\Phi _{0}$ is flux quantum and%
\begin{equation}
\chi ^{\prime \prime }(z)+\frac{\mu }{\hbar ^{2}}\left( 2E_{z}+2e\mathcal{E}%
z-\mu \omega ^{2}z^{2}\right) \chi (z)=0.
\end{equation}%
Consequently, the wave function $g(\rho )$ is required to satisfy the
boundary conditions,\textit{\ i.e.},$\ $\ $g(0)=0$ and $g(\rho \rightarrow
\infty )=0.$\tablenotemark[1]%
\tablenotetext[1]{The solution of
Eq. (7) and Eq. (9) are known in [49].} In order to solve Eq. (7) by means
of the NU method, we introduce the new variable $s=\rho ^{2},$ $\rho \in
(0,\infty )\rightarrow $s$\in (0,\infty )$ which recasts Eq. (7) as in the
following hypergeometric type differential equation:%
\begin{equation}
g^{\prime \prime }(s)+\frac{2}{(2s)}g^{\prime }(s)+\frac{1}{(2s)^{2}}\left(
-\gamma ^{2}s^{2}+\lambda _{1}^{2}s-\beta ^{2}\right) g(s)=0,
\end{equation}%
with 
\begin{subequations}
\begin{equation}
\lambda _{1}^{2}=\frac{2\mu }{\hbar ^{2}}E_{\rho }-\frac{\mu \omega _{c}}{%
\hbar }\left( m+\alpha \right) ,
\end{equation}%
\begin{equation}
\beta ^{2}=\left( m+\alpha \right) ^{2},
\end{equation}%
\begin{equation}
\gamma =\frac{\mu \Omega }{2\hbar },
\end{equation}%
where we have set $g(\rho )\equiv g(s).$ Now, we apply the basic ideas of NU
method [41-45]. Comparing Eq. (10) with the standard form of the
hypergeometric differential equation: 
\end{subequations}
\begin{equation*}
f^{\prime \prime }(s)+\frac{\widetilde{\tau }(s)}{\sigma (s)}f^{\prime }(s)+%
\frac{\widetilde{\sigma }(s)}{\sigma ^{2}(s)}f(s)=0,
\end{equation*}%
gives us the polynomials, 
\begin{equation}
\widetilde{\tau }(s)=2,~~~{\sigma }(s)=2s,~~~\widetilde{\sigma }(s)=-\gamma
^{2}s^{2}+\lambda _{1}^{2}s-\beta ^{2},
\end{equation}%
and further substituting the above polynomials into the expression $\pi (s),$%
\begin{equation*}
\pi (s)=\frac{\sigma ^{\prime }(s)-\widetilde{\tau }(s)}{2}\pm \sqrt{\left( 
\frac{\sigma ^{\prime }(s)-\widetilde{\tau }(s)}{2}\right) ^{2}-\widetilde{%
\sigma }(s)+k{\sigma }(s)},
\end{equation*}%
gives 
\begin{equation}
\pi (s)=\pm \frac{1}{2}\sqrt{\gamma ^{2}s^{2}+(2k-\lambda _{1}^{2})s+\beta
^{2}}.
\end{equation}%
The expression under the square root of the above equation must be the
square of a polynomial of first degree. This is possible only if its
discriminant is zero and the constant parameter $k$ can be determined from
the condition that the expression under the square root has a double zero.
Hence, $k$ is obtained as $k_{+,-}=\lambda ^{2}/2\pm \beta \gamma $. In that
case, it can be written in the four possible forms of $\pi (s)$; 
\begin{equation}
\pi (s)=\left\{ 
\begin{array}{cc}
+\left( \gamma s\pm \beta \right) , & \text{for }k_{+}=\frac{1}{2}\lambda
_{1}^{2}+\beta \gamma , \\ 
-\left( \gamma s\pm \beta \right) , & \text{for }k_{-}=\frac{1}{2}\lambda
_{1}^{2}-\beta \gamma .%
\end{array}%
\right. 
\end{equation}%
One of the four possible forms of $\pi (s)$ must be chosen to obtain an
energy spectrum formula. Therefore, the most suitable form can be
established by the choice:%
\begin{equation*}
\pi (s)=\beta -\gamma s,
\end{equation*}%
for $k_{-}$. The trick in this selection is to find the negative derivative
of $\tau (s)$ given in%
\begin{equation*}
\tau (s)=\widetilde{\tau }(s)+2\pi (s),
\end{equation*}%
which yields 
\begin{equation}
\tau (s)=2\left( 1+\beta \right) -2\gamma s,\text{ }\tau ^{\prime
}(s)=-2\gamma <0~.
\end{equation}%
In this case, it is necessary to use the quantity $\lambdabar _{n}=-n\tau
^{\prime }(s)-\frac{n(n-1)}{2}\sigma ^{\prime \prime }(s)$ to obtain the
eigenvalue equation: 
\begin{equation}
\lambdabar _{n}=2\gamma n,\text{ }n=0,1,2,\ldots 
\end{equation}%
where $n=0,1,2,\ldots $ is the radial quantum number. Another eigenvalue
equation is obtained via the equality $\lambdabar =k_{-}+\pi ^{\prime },$ 
\begin{equation}
\lambdabar =\frac{1}{2}\lambda _{1}^{2}-\gamma \left( \beta +1\right) .
\end{equation}%
Thus to find energy equation, we let $\lambdabar _{n}=\lambdabar $ and the
result obtained will depend on $E_{\rho }$ in the closed form: 
\begin{equation}
\lambda _{1}^{2}=2\left( 2n+1+\beta \right) \gamma ,
\end{equation}%
Upon the substitution of the terms on the right-hand sides of Eqs.
(11a)-(11c) into Eq. (18), we immediately obtain the non-equidistant
magneto-optical energy spectrum for the QDs confinement parabolic potential
as 
\begin{equation}
E_{nm}(\alpha )=E_{\rho }=\frac{1}{2}\hbar \omega _{c}\left( m+\alpha
\right) +\hbar \omega _{c}\left( n+\frac{\left\vert \beta \right\vert +1}{2}%
\right) \sqrt{1+4\left( \frac{\omega }{\omega _{c}}\right) ^{2}},
\end{equation}%
where $\left\vert \beta \right\vert =\left\vert m\right\vert +\alpha >0$ is
an integer$.$ It is apparent from Eq. (19) that the electronic energy levels
are nondegenerate for all $m.$ We have one set of quantum numbers $%
(n,m,\beta )$ for a spinless electron in QDs. Therefore, the energy formula
(19) may be readily used to study the thermodynamic properties of quantum
structures with QDs confined by the harmonic oscillator potential in the
presence and absence of external magnetic field $\left( H\right) $ and AB
flux field $\left( \Phi _{AB}\right) .$

Four special cases are of a particular interest:

\begin{itemize}
\item In the presence of a strong magnetic field; say , $\omega _{c}/\omega
=30$ [48]$,$ then$\ \Omega \rightarrow \omega _{c}\gg \omega ,$ then $%
E_{nm}(\alpha )=\hbar \omega _{c}\left[ n+\alpha +\frac{1}{2}(m+\left\vert
m\right\vert +1)\right] $ which is the formula in the presence of magnetic $%
\left( H\right) $ and AB flux $\left( \Phi _{AB}\right) $ fields$.$
Meanwhile, in the presence of a weak magnetic field; say , $\omega
_{c}/\omega =3$ [48], then we can resort to Eq. (19).

\item If we set $\alpha =0,$ i.e., in the absence of AB flux field and
presence of strong magnetic field, we find $E_{nm}=\hbar \omega _{c}\left[ n+%
\frac{1}{2}(\left\vert m\right\vert +m+1)\right] .$

\item In the absence of magnetic field ($\omega _{c}=0$) and an AB flux
field ($\alpha =0$), we find $E_{nm}=\hbar \omega \left( 2n+\left\vert
m\right\vert +1\right) .$

\item The case $m=0$ is simply for harmonic oscillator energy spectrum,
i.e., $E_{n}=\hbar \omega \left( 2n+1\right) $.
\end{itemize}

Next, we need to calculate the corresponding wave function for the
confinement potential model. We find the first part of the wave function by 
\begin{equation}
\phi _{m}(s)=\exp \left( \int \frac{\pi (s)}{\sigma (s)}ds\right)
=s^{\left\vert \beta \right\vert /2}e^{-\gamma s/2}.
\end{equation}%
Then, the weight function defined by 
\begin{equation}
\rho (s)=\frac{1}{\sigma (s)}\exp \left( \int \frac{\tau (s)}{\sigma (s)}%
ds\right) =s^{\left\vert \beta \right\vert }e^{-\gamma s},
\end{equation}%
which gives the second part of the wave function (Rodrigues formula) given by%
\begin{equation*}
y_{nm}(s)=y^{(n)}(s)=y(s)=\frac{B_{n}}{\rho (s)}\frac{d^{n}}{dr^{n}}\left[
\sigma ^{n}(s)\rho (s)\right] ,
\end{equation*}%
or alternatively%
\begin{equation}
y_{nm}(s)\sim s^{-\left\vert \beta \right\vert }e^{\gamma s}\frac{d^{n_{r}}}{%
ds^{n_{r}}}\left( s^{n+\left\vert \beta \right\vert }e^{-\gamma s}\right)
\sim L_{n}^{\left( \left\vert \beta \right\vert \right) }\left( \gamma
s\right) ,
\end{equation}%
where $L_{a}^{\left( b\right) }\left( x\right) =\frac{\left( a+b\right) !}{%
a!b!}F\left( a,b+1;x\right) $ is the associated Laguerre polynomial and $%
F(a,b+1;x)$ is the confluent hypergeometric function. With the formula $%
g(s)=\phi _{m}(s)y_{nm}(s),$ we may write the radial wave function as%
\begin{equation}
g(\rho )=C_{n,m}\rho ^{\left\vert m\right\vert +\alpha }e^{-\gamma \rho
^{2}/2}F\left( -n,\left\vert m\right\vert +\alpha +1;\gamma \rho ^{2}\right)
,
\end{equation}%
and finally the total wave function (6) reads%
\begin{equation*}
R_{n,m}(\rho ,\varphi )=\frac{1}{a^{\left( 1+\left\vert m\right\vert +\alpha
\right) }}\left[ \frac{n!}{\pi 2^{\left\vert m\right\vert +\alpha +1}\left(
\left\vert m\right\vert +\alpha +n\right) !}\right] ^{1/2}\rho ^{\left\vert
m\right\vert +\alpha }e^{-\rho ^{2}/4a^{2}}L_{n}^{\left( \left\vert
m\right\vert +\alpha \right) }\left( \rho ^{2}/2a^{2}\right) e^{im\varphi },
\end{equation*}%
\begin{equation}
=\frac{a^{-\left( 1+\left\vert m\right\vert +\alpha \right) }}{\left(
\left\vert m\right\vert +\alpha \right) !}\left[ \frac{\left( \left\vert
m\right\vert +\alpha +n\right) !}{\pi 2^{\left\vert m\right\vert +\alpha
+1}n!}\right] ^{1/2}\rho ^{\left\vert m\right\vert +\alpha }e^{-\rho
^{2}/4a^{2}}F\left( -n,\left\vert m\right\vert +\alpha +1;\rho
^{2}/2a^{2}\right) e^{im\varphi },
\end{equation}%
where $a=\sqrt{\hbar /\mu \Omega }$ is the effective length scale. The
energy levels (19) with $\alpha =0$ (i.e., $\Phi _{AB}=0$) becomes 
\begin{equation}
E_{nm}=\frac{1}{2}\hbar \omega _{c}m+\hbar \Omega \left( n+\frac{\left\vert
m\right\vert +1}{2}\right) ,
\end{equation}%
and the wave function (24) becomes 
\begin{equation}
R_{n,m}(\rho ,\varphi )=\frac{a^{-\left( 1+\left\vert m\right\vert \right) }%
}{\left\vert m\right\vert !}\left[ \frac{\left( n_{\rho }+\left\vert
m\right\vert \right) !}{\pi 2^{\left\vert m\right\vert +1}n_{\rho }!}\right]
^{1/2}e^{im\varphi }\rho ^{\left\vert m\right\vert }e^{-\rho
^{2}/4a^{2}}F\left( -n,\left\vert m\right\vert +1;\rho ^{2}/2a^{2}\right) .
\end{equation}%
which are identical to Eqs. (10) and (8) in Ref. [1], respectively. On the
other hand, Eq. (9) can be recasted in the form: 
\begin{subequations}
\begin{equation}
\chi ^{\prime \prime }(z)+\left( -\delta ^{2}z^{2}+\eta ^{2}z-\epsilon
^{2}\right) \chi (z)=0,
\end{equation}%
\begin{equation}
\delta =\frac{\mu \omega }{\hbar },\text{ }\eta =\sqrt{\frac{2\mu e\mathcal{E%
}}{\hbar ^{2}}},\text{ }\epsilon =\sqrt{-\frac{2\mu E_{z}}{\hbar ^{2}}},%
\text{ }E_{z}<0.
\end{equation}%
We follow the same procedures of solution by writting 
\end{subequations}
\begin{equation}
\widetilde{\tau }(z)=0,~~~{\sigma }(z)=1,~~~\widetilde{\sigma }(z)=-\delta
^{2}z^{2}+\eta ^{2}z-\epsilon ^{2}.
\end{equation}%
In the present case, the polynomial $\pi (z)$ is obtained as 
\begin{equation}
\pi (z)=\pm \frac{1}{2}\sqrt{\delta ^{2}z^{2}-\eta ^{2}z+\epsilon ^{2}+k},
\end{equation}%
and thus the two possible forms of $\pi (z)$ are 
\begin{equation}
\pi (z)=\left\{ 
\begin{array}{cc}
+\delta \left( z-\eta ^{2}/2\delta ^{2}\right) , & \text{for }k=\eta
^{4}/4\delta ^{2}-\epsilon ^{2}, \\ 
-\delta \left( z-\eta ^{2}/2\delta ^{2}\right) , & \text{for }k=\eta
^{4}/4\delta ^{2}-\epsilon ^{2}.%
\end{array}%
\right. 
\end{equation}%
Therefore, the most suitable form can be established by the choice:%
\begin{equation*}
\pi (z)=-\delta \left( z-\eta ^{2}/2\delta ^{2}\right) ,
\end{equation*}%
and $\tau (z)$ is consequently found as 
\begin{equation}
\tau (z)=-2\delta z+\eta ^{2}/\delta .
\end{equation}%
A new eigenvalue equation becomes 
\begin{equation}
\lambdabar _{n_{z}}=2\delta n_{z},\text{ }n_{z}=0,1,2,\ldots 
\end{equation}%
where $n_{z}$ is the quantum number and another eigenvalue equation is
obtained as 
\begin{equation}
\lambdabar =\frac{\eta ^{4}}{4\delta ^{2}}-\epsilon ^{2}-\delta .
\end{equation}%
Hence, the energy formula reads as 
\begin{equation}
E_{z}=\hbar \omega \left( n_{z}+\frac{1}{2}\right) -\frac{e^{2}\mathcal{E}%
^{2}}{2\mu \omega ^{2}}.
\end{equation}%
Next, we calculate the wave function $\chi (z)$. The first part of the wave
function is%
\begin{equation}
\phi (z)=e^{-\delta \left( z-\eta ^{2}/2\delta ^{2}\right) ^{2}/2+\eta
^{4}/8\delta ^{3}},
\end{equation}%
and the weight function\ is 
\begin{equation}
\rho (z)=e^{-\delta \left( z-\eta ^{2}/2\delta ^{2}\right) ^{2}+\eta
^{4}/4\delta ^{3}},
\end{equation}%
which gives the second part of the wave function: 
\begin{equation*}
\chi _{n_{z}}(z)\sim (-1)^{n_{z}}e^{-\delta \left( z-\eta ^{2}/2\delta
^{2}\right) ^{2}/2}H_{n_{z}}\left[ \sqrt{\delta }\left( z-\eta ^{2}/2\delta
^{2}\right) \right] 
\end{equation*}%
\begin{equation}
=\frac{1}{\sqrt{2^{n_{z}}n_{z}!}}\left( \frac{\mu \omega }{\hbar \pi }%
\right) ^{1/4}e^{-\left( \mu \omega /\hbar \right) \left( z-e\mathcal{E}/\mu
\omega ^{2}\right) ^{2}/2}H_{n_{z}}\left[ \sqrt{\mu \omega /\hbar }\left( z-e%
\mathcal{E}/\mu \omega ^{2}\right) \right] ,
\end{equation}%
where $H_{n_{z}}(x)$ is the Hermite polynomial. As to electronic energy
levels, it is the sum of expressions (19) and (34):%
\begin{equation*}
E_{n,n_{z},m}(\alpha ,\omega ,\omega _{c},\mathcal{E})=\hbar \omega \left[
\left( n+\frac{\left\vert m\right\vert +\alpha +1}{2}\right) \sqrt{\left( 
\frac{\omega _{c}}{\omega }\right) ^{2}+4}+\frac{m+\alpha }{2}\left( \frac{%
\omega _{c}}{\omega }\right) +n_{z}+\frac{1}{2}\right] 
\end{equation*}%
\begin{equation}
-\frac{e^{2}\mathcal{E}^{2}}{2\mu \omega ^{2}},
\end{equation}%
where $\hbar \omega _{c}=0.11571589$ $H$ $(meV)$ and $H$ is to be in units
of Tesla. Additionally, the term $\alpha =\Phi _{AB}/\Phi _{0}$ reflects the
dependence of the electronic levels on the AB flux $\Phi _{AB}$ where we
take $\alpha =6.$

As for the wave functions, it is taken as the product of (24) and (37):%
\begin{equation*}
\psi (\rho ,\varphi ,z)=\frac{a^{-\left( 1+\left\vert \beta \right\vert
\right) }}{\left\vert \beta \right\vert !}\left[ \frac{\left( \left\vert
\beta \right\vert +n_{\rho }\right) !}{\pi 2^{\left\vert \beta \right\vert
+1}n_{\rho }!}\right] ^{1/2}e^{im\varphi }\rho ^{\left\vert \beta
\right\vert }e^{-\rho ^{2}/4a^{2}}F\left( -n_{\rho },\left\vert \beta
\right\vert +1;\rho ^{2}/2a^{2}\right) 
\end{equation*}%
\begin{equation}
\times \frac{\left( b\right) ^{-1/2}}{\left( \pi \right) ^{1/4}\sqrt{%
2^{n_{zz}}n_{z}!}}e^{-\left( z-e\mathcal{E}/\mu \omega ^{2}\right)
^{2}/2b^{2}}H_{n_{z}}\left[ \left( z-e\mathcal{E}/\mu \omega ^{2}\right) /b%
\right] ,
\end{equation}%
where $b=\sqrt{\hbar /\mu \omega }.$ Expressions (19) and (39) obtained
above for a charge carrier energy spectrum and wave functions are identical
to Eqs. (10) and (11) in Ref. [1] and Eqs. (4) and (3) in Ref. [50],
respectively, in the case when $\alpha =0.$

To show the behaviour of the energy formula (38), we follow Ref. [51] in
plotting it versus $\omega _{c}/\omega $ under the influence of magnetic and
AB flux fields. In figure 1, we plot the eigenenergies (in units $\hbar
\omega )$ versus the ratio $\omega _{c}/\omega $ (a) for various AB field, $%
\alpha $ (b) for various magnetic quantum number $m$ and (c) for various
quantum number $n_{z}.$ As shown in figure 1a, the ground state $n=n_{z}=0$
(singlet state, $m=0$) under the influence of the AB field leads to the
phase transition to the high-lying states $n>0.$ The family of states for
various values of $\alpha =0,1,2,3,4$ are non-linear in the presence of weak
magnetic field $0<\omega _{c}/\omega <5.$ In figure1b, the energy
eigenvalues (in units $\hbar \omega )$ is found increasing with the
increasing magnetic quantum number $m$ when $\omega _{c}>\omega .$ However,
when $\omega _{c}<\omega ,$ we notice the crossing between $m=1$ and $m=-1$
states. As shown in figure 1c, the ground state $n=m=0$ leads to the phase
transition to the high-lying states $n>0$ when the quantum number $n_{z}$ is
increasing. As $n_{z}$ is increasing we have a family of states for various $%
n_{z}=0,1,2,3,4.$

\subsection{Interband light absorption coefficient}

Expressions (38) and (39), obtained above for electronic energy spectrum and
corresponding wave functions in cylindrical QDs influenced by external
uniform electrical and magnetic fields along with an AB flux field, allow us
to calculate the direct interband light absorption coefficient $K(\overline{%
\omega })$ in the present system and the threshold frequency of absorption.
In case of strong size quantization in which it is possible to neglect the
electron-hole interactions. According to Ref. [52], the light absorption
coefficient is 
\begin{equation*}
K(\overline{\omega })=N\dsum\limits_{n,m,n_{z}}\dsum\limits_{n^{\prime
},m^{\prime },n_{z}^{\prime }}\left\vert \dint \psi _{n,m,n_{z}}^{e}(\rho
,\varphi ,z)\psi _{n^{\prime },m^{\prime },n_{z}^{\prime }}^{h}(\rho
,\varphi ,z)\rho d\rho d\varphi dz\right\vert ^{2}\delta \left( \Delta
-E_{n,m,n_{z}}^{e}-E_{n^{\prime },m^{\prime },n_{z}^{\prime }}^{h}\right)
\end{equation*}%
\begin{equation*}
=N\dsum\limits_{n,m,n_{z}}\dsum\limits_{n^{\prime },m,n_{z}^{\prime }}\frac{%
4\left( \frac{1}{4a_{e}^{2}a_{h}^{2}}\right) ^{1+\left\vert \beta
\right\vert }}{\left( \left\vert \beta \right\vert !\right) ^{4}}\frac{%
\left( \left\vert \beta \right\vert +n\right) !\left( \left\vert \beta
\right\vert +n^{\prime }\right) !}{2^{n_{z}+n^{\prime }{}_{z}}n!n^{\prime }!}%
\frac{\left( \frac{\mu _{e}\omega }{\hbar }\right) ^{1/2}\left( \frac{\mu
_{h}\omega }{\hbar }\right) ^{1/2}}{\pi n_{z}!n_{z}^{\prime }!}
\end{equation*}%
\begin{equation*}
\times \left\vert \dint_{0}^{\infty }\rho d\rho \rho ^{2\left\vert \beta
\right\vert }e^{-\left( 1/4a_{e}^{2}+1/4a_{h}^{2}\right) \rho ^{2}}F\left(
-n,\left\vert \beta \right\vert +1;\rho ^{2}/2a_{e}^{2}\right) F\left(
-n^{\prime },\left\vert \beta \right\vert +1;\rho ^{2}/2a_{h}^{2}\right)
\right\vert ^{2}
\end{equation*}%
\begin{equation}
\times I_{nn^{\prime }}^{2}\delta \left( \Delta
-E_{n,m,n_{z}}^{e}-E_{n^{\prime },m,n_{z}^{\prime }}^{h}\right) ,
\end{equation}%
where $\Delta =\hbar \overline{\omega }-\varepsilon _{g},$ $\varepsilon _{g}$
is the width of forbidden energy gap, $\overline{\omega }$ is the frequency
of incident light, $N$ is a quantity proportional to the square of dipole
moment matrix element modulus, $\psi ^{e(h)}$ is the wave function of the
electron (hole) and $E^{e(h)}$ is the corresponding energy of the electron
(hole).

Now, we use the integrals [53]%
\begin{equation}
\dint_{0}^{2\pi }e^{i\left( m+m^{\prime }\right) \phi }d\phi =\left\{ 
\begin{array}{ccc}
2\pi  & \text{if} & m=-m^{\prime }, \\ 
0 & \text{if} & m\neq -m^{\prime },%
\end{array}%
\right. 
\end{equation}%
\begin{equation*}
\dint_{0}^{\infty }e^{-\kappa x}x^{\lambda -1}F\left( -n,\lambda ;qx\right)
F\left( -n^{\prime },\lambda ;q^{\prime }x\right) dx=\Gamma (\lambda )\kappa
^{n+n^{\prime }-\lambda }\left( \kappa -q\right) ^{-n}\left( \kappa
-q^{\prime }\right) ^{-n^{\prime }}
\end{equation*}%
\begin{equation}
\times 
\begin{array}{c}
_{2}F_{1}%
\end{array}%
\left( n,n^{\prime },\lambda ;\frac{qq^{\prime }}{\left( \kappa -q\right)
\left( \kappa -q^{\prime }\right) }\right) ,
\end{equation}%
and%
\begin{equation*}
I_{nn^{\prime }}=\dint_{-\infty }^{\infty }e^{-\left( \frac{\mu _{e}\omega
_{e}}{2\hbar }\right) \left( z-\frac{e\mathcal{E}}{\mu _{e}\omega _{e}^{2}}%
\right) ^{2}}e^{-\left( \frac{\mu _{h}\omega _{h}}{2\hbar }\right) \left( z+%
\frac{e\mathcal{E}}{\mu _{h}\omega _{h}^{2}}\right) ^{2}}
\end{equation*}%
\begin{equation*}
\times H_{n}\left[ \left( \frac{\mu _{e}\omega _{e}}{\hbar }\right) \left( z-%
\frac{e\mathcal{E}}{\mu _{e}\omega _{e}^{2}}\right) \right] H_{n^{\prime }}%
\left[ \left( \frac{\mu _{h}\omega _{h}}{\hbar }\right) \left( z+\frac{e%
\mathcal{E}}{\mu _{h}\omega _{h}^{2}}\right) \right] dz,
\end{equation*}%
where $\Gamma (x)$ is the Euler-Gamma function and $%
\begin{array}{c}
_{2}F_{1}%
\end{array}%
\left( a,b,c;z\right) $ is the hypergeometric function, to calculate the
light absorption coefficient:%
\begin{equation}
K(\overline{\omega })=N\dsum\limits_{n,m,\beta }\dsum\limits_{n^{\prime
},m^{\prime },\beta ^{\prime }}P_{n,n^{\prime }}^{\beta }Q_{n,n^{\prime
}}^{\beta }\delta \left( \Delta -E_{n,m,\beta }^{e}-E_{n^{\prime },m^{\prime
},\beta ^{\prime }}^{h}\right) ,
\end{equation}%
where%
\begin{equation}
P_{n,n^{\prime }}^{\beta }=\frac{1}{\left( \left\vert \beta \right\vert
!\right) ^{2}}\left( \frac{\mu _{e}\omega }{\hbar }\right) ^{1/2}\left( 
\frac{\mu _{h}\omega }{\hbar }\right) ^{1/2}\left( \gamma \gamma ^{\prime
}\right) ^{\left\vert \beta \right\vert +1}\left( \frac{\gamma +\gamma
^{\prime }}{\gamma -\gamma ^{\prime }}\right) ^{2\left( n+n^{\prime }\right)
}\frac{\left( n+\left\vert \beta \right\vert \right) !\left( n^{\prime
}+\left\vert \beta \right\vert \right) !}{n!n^{\prime }!\pi
n_{z}!n_{z}^{\prime }!2^{n_{z}+n^{\prime }{}_{z}}},
\end{equation}%
and%
\begin{equation}
Q_{n,n^{\prime }}^{\beta }=\left[ \left( \frac{2}{\gamma +\gamma ^{\prime }}%
\right) ^{\left\vert \beta \right\vert +1}%
\begin{array}{c}
_{2}F_{1}%
\end{array}%
\left( n,n^{\prime },\left\vert \beta \right\vert +1;-\frac{4\gamma \gamma
^{\prime }}{\left( \gamma -\gamma ^{\prime }\right) ^{2}}\right) \right]
^{2},\text{ }\gamma =\frac{1}{2a_{e}^{2}},\text{ }\gamma ^{\prime }=\frac{1}{%
2a_{h}^{2}}.
\end{equation}%
Using Eq. (38), we find the threshold frequency value of absorption as%
\begin{equation*}
\hbar \overline{\omega }=\varepsilon _{g}+\hbar \left( n+\frac{\left\vert
m\right\vert +\alpha +1}{2}\right) \left( \frac{e^{2}H^{2}}{\mu ^{2}c^{2}}%
+4\omega ^{2}\right) ^{1/2}+\hbar \frac{eH}{\mu c}\left( \frac{m+\alpha }{2}%
\right) +\hbar \omega \left( n_{z}+\frac{1}{2}\right) -\frac{e^{2}\mathcal{E}%
^{2}}{2\mu \omega ^{2}}
\end{equation*}%
\begin{equation}
+\hbar \left( n^{\prime }+\frac{\left\vert m^{\prime }\right\vert +\alpha +1%
}{2}\right) \left( \frac{e^{2}H^{2}}{\mu ^{\prime 2}c^{2}}+4\omega
^{2}\right) ^{1/2}+\hbar \frac{eH}{\mu ^{\prime }c}\left( \frac{m^{\prime
}+\alpha }{2}\right) +\hbar \omega \left( n_{z}^{\prime }+\frac{1}{2}\right)
-\frac{e^{2}\mathcal{E}^{2}}{2\mu ^{\prime }\omega ^{2}}.
\end{equation}%
For ground state, we set $n=m=0$ in the above expression to obtain the
threshold frequency of absorption:%
\begin{equation*}
\overline{\omega }_{00}=\frac{\varepsilon _{g}}{\hbar }+\left( \frac{\alpha
+1}{2}\right) \left( \frac{e^{2}H^{2}}{\mu ^{2}c^{2}}+4\omega ^{2}\right)
^{1/2}-\frac{e^{2}\mathcal{E}^{2}}{2\hbar \omega ^{2}}\left( \frac{1}{\mu }+%
\frac{1}{\mu ^{\prime }}\right) 
\end{equation*}%
\begin{equation}
+\left( \frac{\alpha +1}{2}\right) \left( \frac{e^{2}H^{2}}{\mu ^{\prime
2}c^{2}}+4\omega ^{2}\right) ^{1/2}+\frac{eH\alpha }{2c}\left( \frac{1}{\mu }%
+\frac{1}{\mu ^{\prime }}\right) +\omega .
\end{equation}%
We follow Ref. [54] in plotting the threshold frequency of absorption $%
\overline{\omega }_{00}$ (in units of $\varepsilon _{g}$ ) versus the
magnetic field strength $H$ and quantum dot size considering various AB
magnetic flux values $\alpha =0,1,2,3.$ In figure 2, we plot the variations
of threshold frequency of absorption $\overline{\omega }_{00}$ (in units of $%
\varepsilon _{g}$ ) as a function of applied (a) large magnetic field and
(b) small magnetic field in unit of $h=\left( e\hbar H/\mu c\varepsilon
_{g}\right) $ with $\rho =89.53$. It is seen from figure 2a (figure 2b) that
the dependence of $\overline{\omega }_{00}$ on $H$ is linear (nonlinear) for
large (small) applied magnetic fields. The main feature in the application
of the AB flux field leads to a family of the phase transition for the
ground state $n=0,$ mainly $\alpha =0,1,2,3$ leads to a phase transitions
for the high-lying states $n>0.$ In figure 3, we plot the threshold
frequency of absorption $\overline{\omega }_{00}$ (in units of $\varepsilon
_{g}$ ) as a function of quantum dot size (in unit of $\rho =\sqrt{%
\varepsilon _{g}/\omega \hbar }=\sqrt{\mu \varepsilon _{g}/\hbar ^{2}}r_{0}$%
) (see Eq. (3)) with $h=0.062$. It is seen in figure 3 that the threshold
frequency of absorption decreases when the quantum dot size increasing. The
application of AB flux field $\Phi _{AB}$ generates a family of state
transitions for $\alpha =\Phi _{AB}/\Phi _{0}=0,1,2,3.$

\section{Concluding Remarks}

In this work, we have obtained the bound state solutions of the Schr\"{o}%
dinger spinless particle in QDs confined to non-relativistic harmonic
oscillator in presence of electrical, magnetic and AB flux fields. The
electron (hole) energy spectrum and the corresponding wave functions are
used to calculate the the interband light absorption coefficient and the
threshold frequency of absorption. Also, the energy spectrum of the electron
may be used to study the thermodynamic properties of quantum structures with
dot in electrical, magnetic and AB flux fields. The electronic energy levels
make a shift under the effect of an external electrical field by an amount $%
\Delta E=-e^{2}\mathcal{E}^{2}/$ ($2\mu \omega ^{2}$)$.$ It explains the
Stark splitting quadratic dependence on $\overrightarrow{\mathcal{E}}$. The
energy levels in the presence of external electrical field of different
strengths are nondegenerate. The threshold frequency of absorption $%
\overline{\omega }_{00}$ rises on the field $\overrightarrow{\mathcal{E}}$
by quadratic law and has also more complicated dependence on the magnetic
field $\overrightarrow{H}.$ Further, it is noticed that the spinless
particle (electron) is localized along the z-axis inside the QDs.

In the quantum mechanics there is a relevant relationship between 2D and 3D
harmonic oscillator [35] in Schr\"{o}dinger theory. We find that there are
corresponding relationship between 2D and 3D harmonic oscillators, $\rho
\leftrightarrow r$ and $\left\vert m\right\vert \leftrightarrow l+1/2.$

\acknowledgments We acknowledge the partial support provided by the
Scientific and Technological Research Council of Turkey.\newpage

{\normalsize 
}

\bigskip \baselineskip= 2\baselineskip
\bigskip \newpage

\bigskip

\bigskip {\normalsize 
}

\baselineskip= 2\baselineskip
\FRAME{ftbpFO}{0.0277in}{0.0277in}{0pt}{\Qct{Eigenenergies (in units $\hbar 
\protect\omega )$ versus the ratio $\protect\omega _{c}/\protect\omega $ for
a) various AB field $\protect\alpha $, b) various magnetic quantum number $m$
and c) various quantum number $n_{z}$.}}{}{Figure 1}{}

\bigskip \FRAME{ftbpFO}{0.0277in}{0.0277in}{0pt}{\Qct{The variations of
threshold frequency of absorption $\overline{\protect\omega }_{00}$ (in
units of $\protect\varepsilon _{g}$ ) as a function of applied (a) large
magnetic field and (b) small magnetic field (in unit of $h$).}}{}{Figure 2}{}

\FRAME{ftbpFO}{0.0277in}{0.0277in}{0pt}{\Qct{The variations of the threshold
frequency of absorption $\overline{\protect\omega }_{00}$ (in units of $%
\protect\varepsilon _{g}$ ) as a function of quantum dot size (in unit of $%
\protect\rho $).}}{}{Figure 3}{}

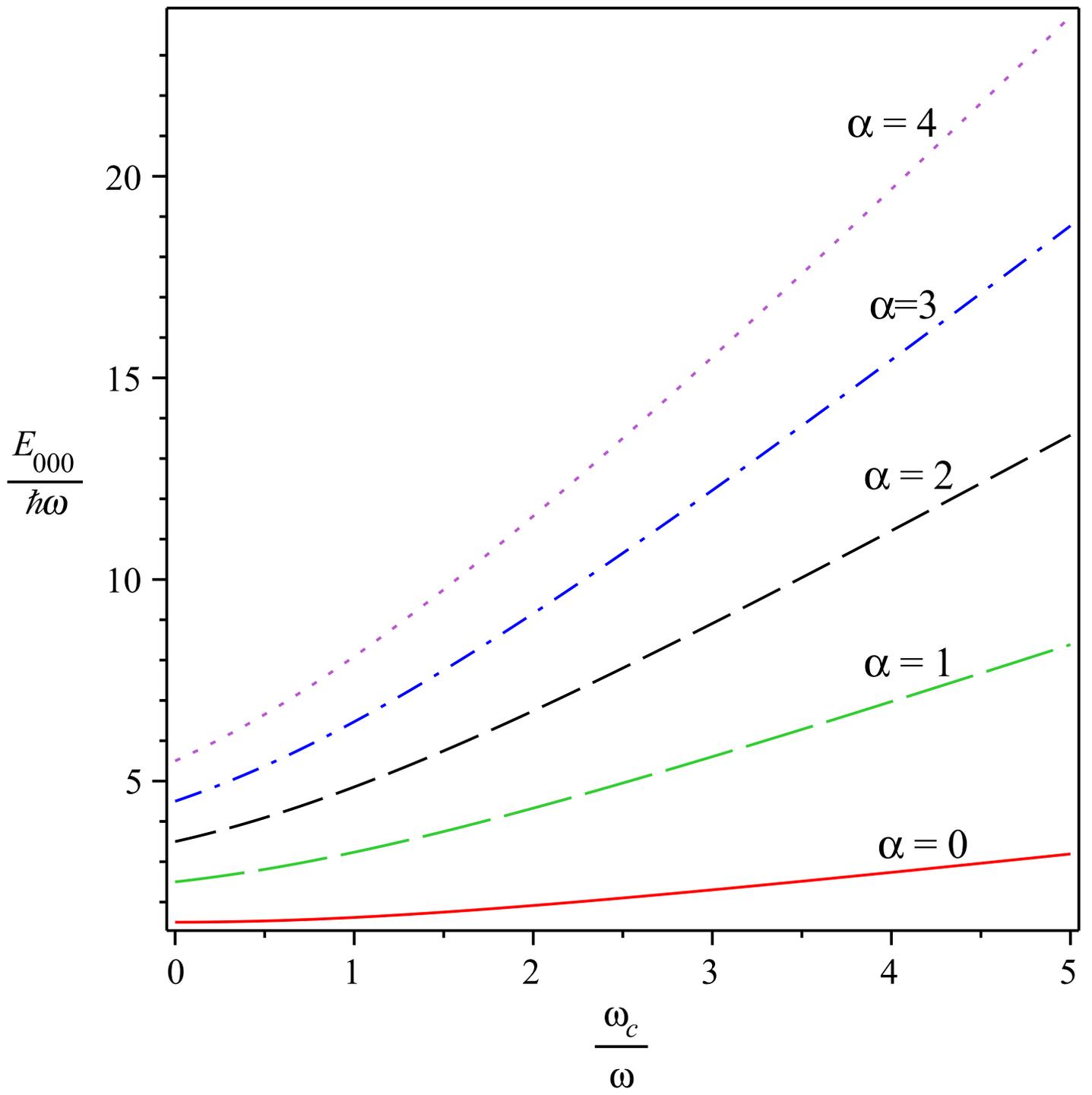
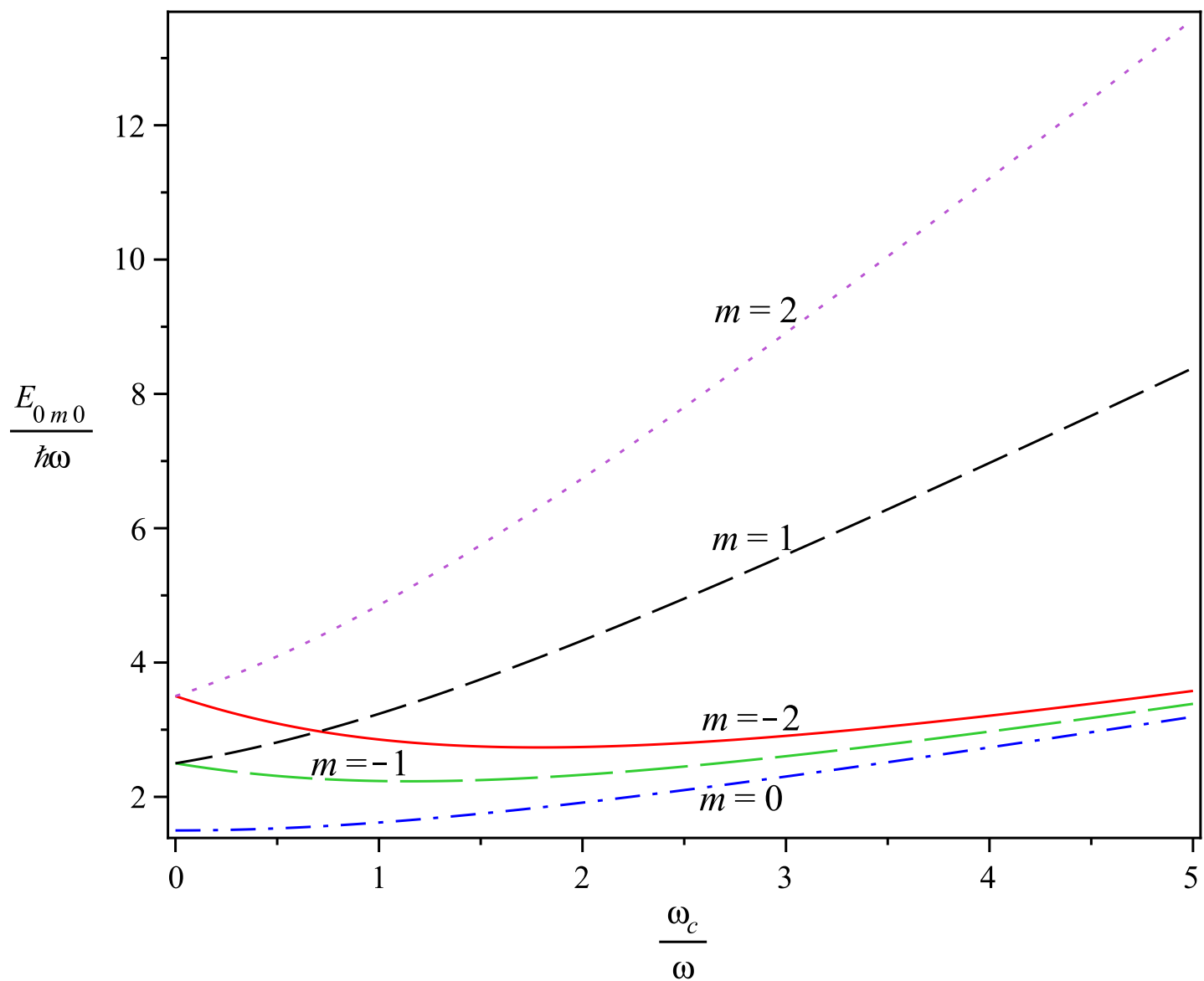
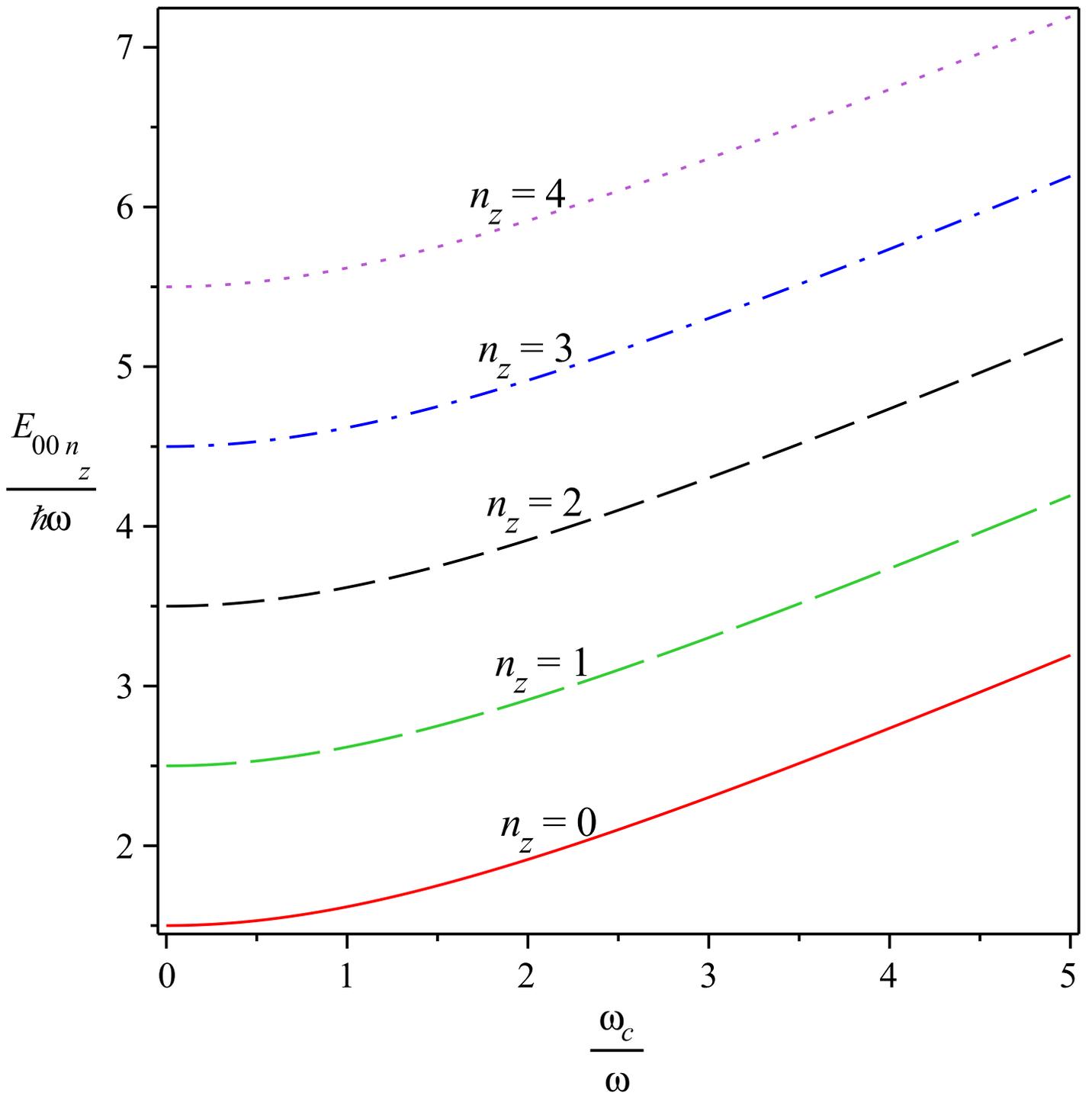
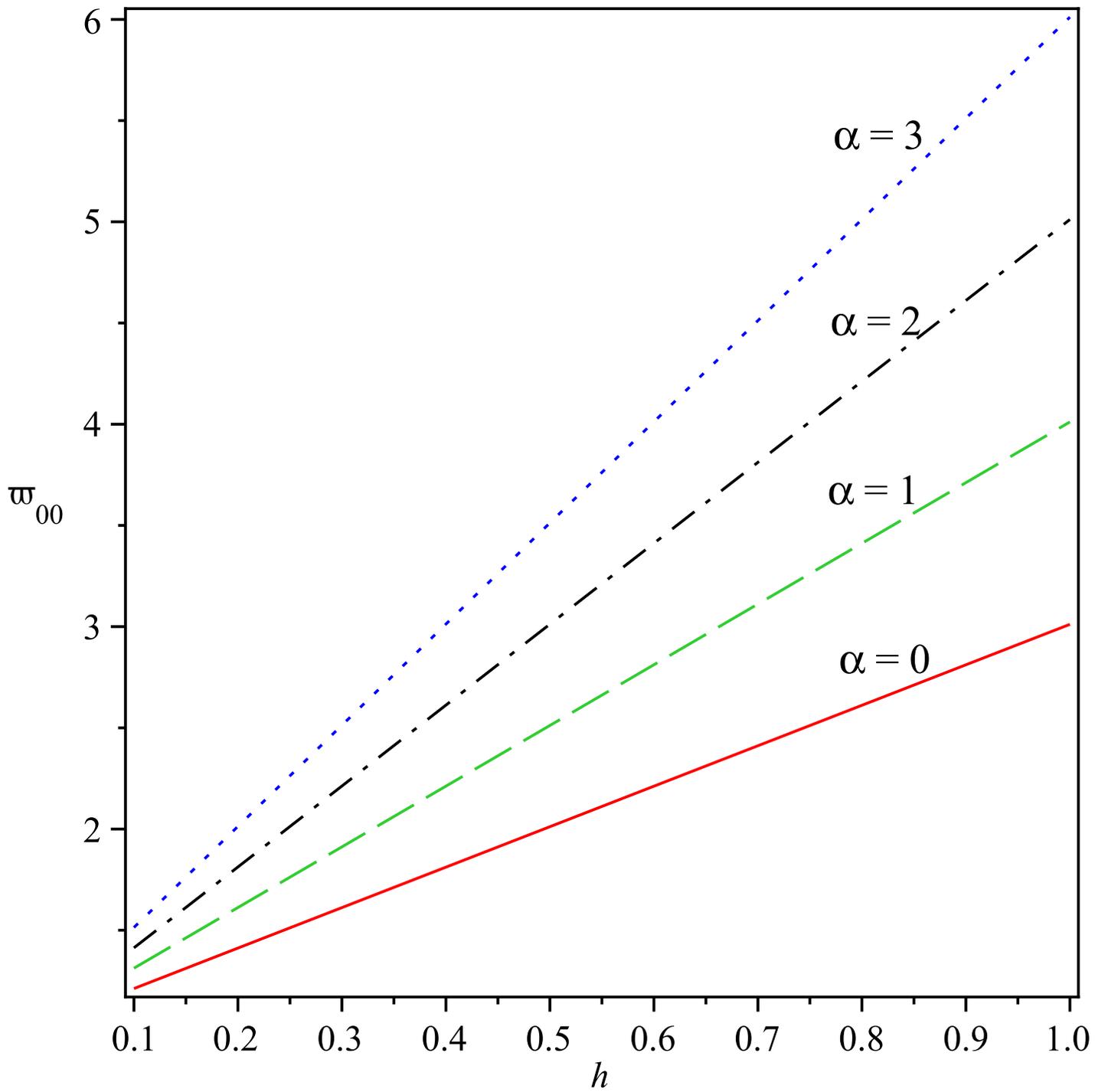
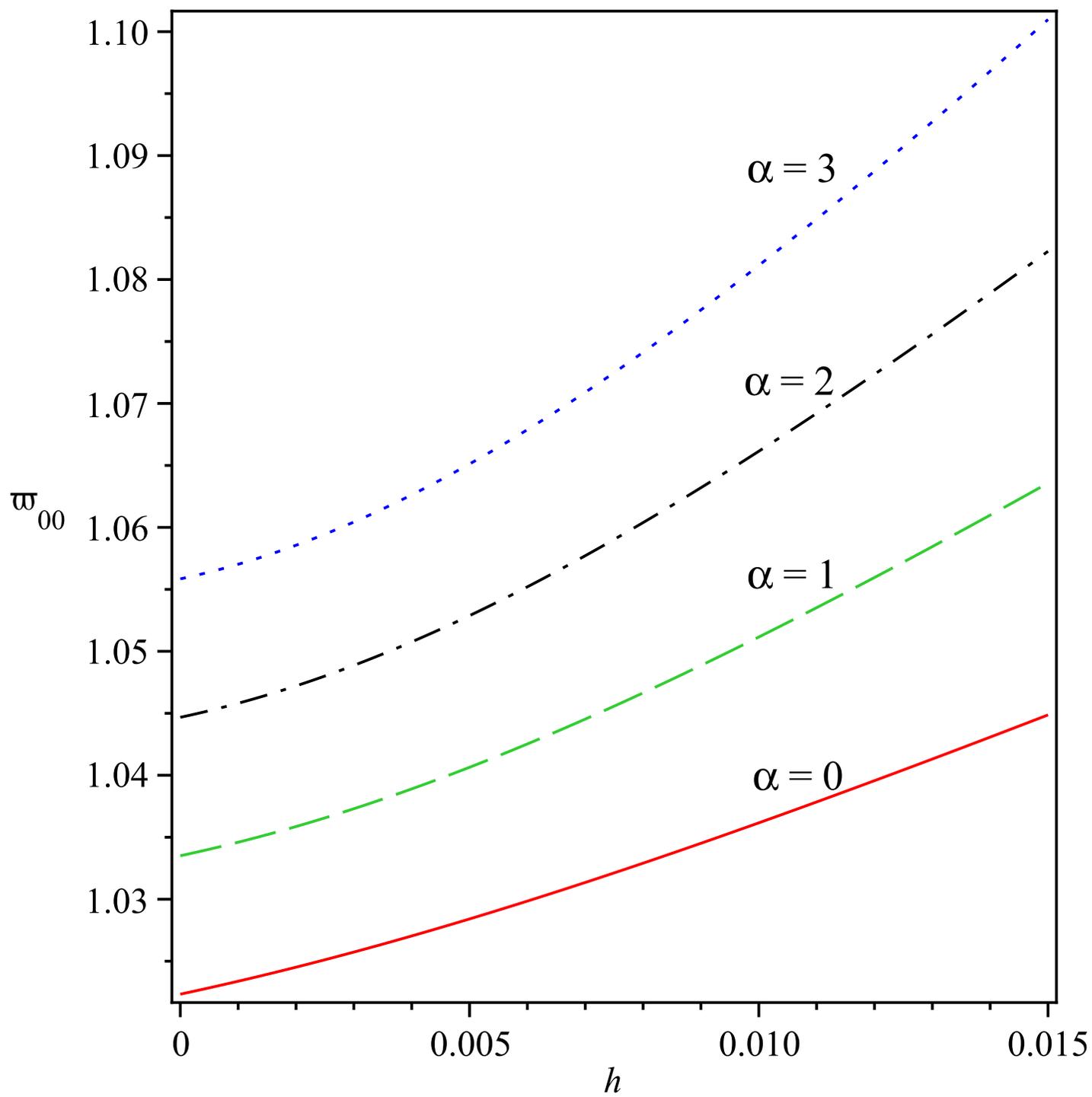
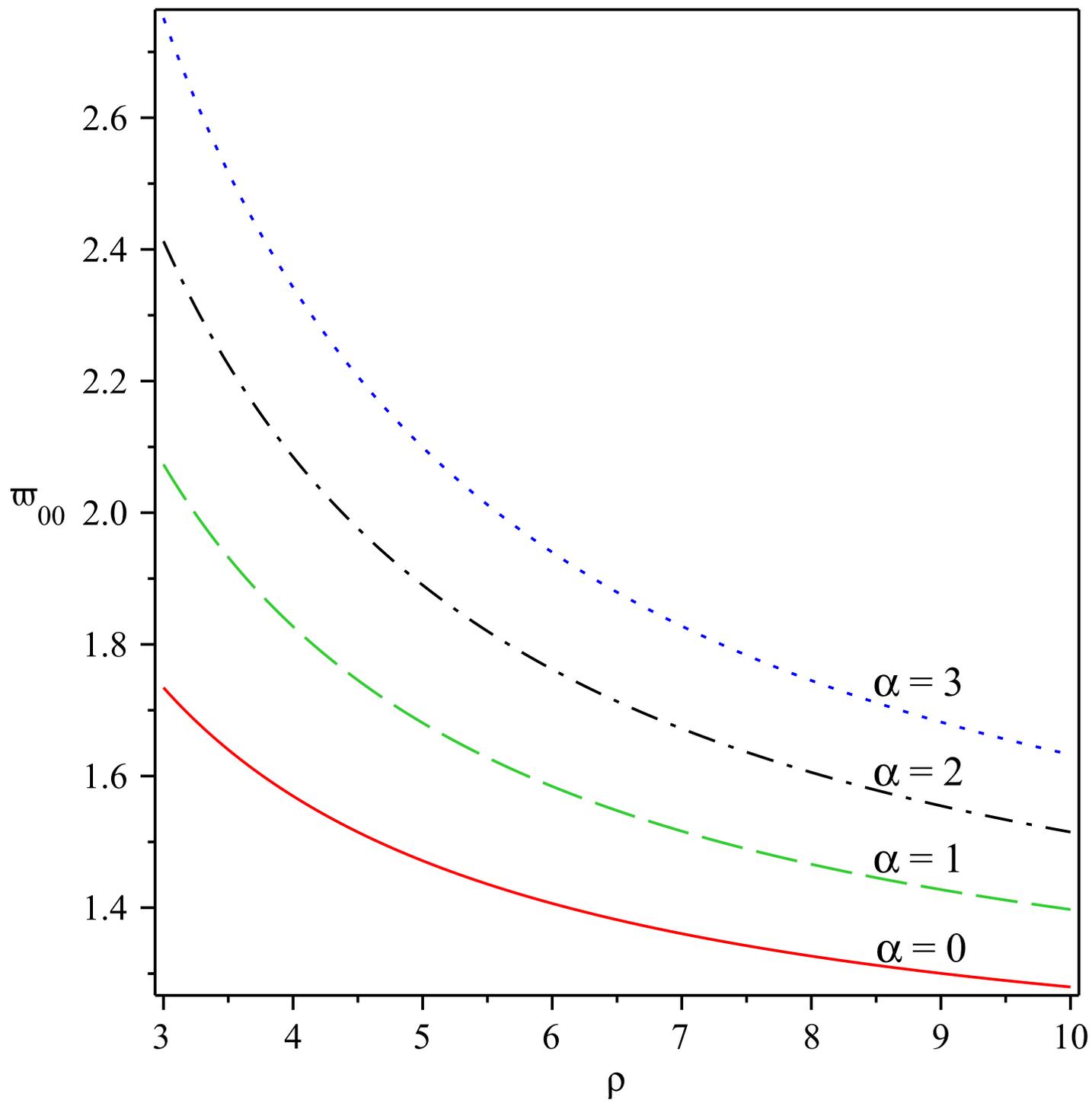
\end{document}